\documentclass{iau} 
\usepackage{graphicx}

\newcommand{\fion}[2]{\mbox{\rm [{#1}\thinspace{\footnotesize {#2}}]}} 
\def\gapeq{\mathrel{\hbox{\rlap{\hbox{\lower4pt\hbox{$\sim$}}}\hbox{$>$}}}}
\newcommand{\HI}{\mbox {H\thinspace{\footnotesize I}}}
\newcommand{\HII}{\mbox {H\thinspace{\footnotesize II}}}
\newcommand{\HIPASS}{{\sc HiPASS}}
\newcommand{\Hline}[1]{\mbox{H{\footnotesize {#1}}}}
\newcommand{\Halpha}{\Hline{\mbox{$\alpha$}}}
\def\lapeq{\mathrel{\hbox{\rlap{\hbox{\lower4pt\hbox{$\sim$}}}\hbox{$<$}}}}
\newcommand{\Msun}{\mbox{${\cal M}_\odot$}}
\newcommand{\Vmax}{\mbox{$V_{\rm max}$}}
\title[Outer Disk Star Formation in HI selected Galaxies] 
{Outer Disk Star Formation in HI selected Galaxies}

\author[G.R.\ Meurer]   
{G.R.\ Meurer$^1$}

\affiliation{$^1$International Centre for Radio Astronomy Research\\email: {\tt gerhardt.meurer@icrar.org}}

\pubyear{2016}
\volume{321}  
\jname{Outskirts of Galaxies}
\editors{A.\ Gil de Paz, J.\ Knapen, J.\ Lee, eds.}
\begin{document}

\maketitle

\begin{abstract}
The HI in galaxies often extends past their conventionally defined optical extent. I report results from our team which has been probing low intensity star formation in outer disks using imaging in \Halpha\ and ultraviolet. Using a sample of hundreds of HI selected galaxies, we confirm that outer disk HII regions and extended UV disks are common. Hence outer disks are not dormant but are dimly forming stars.  Although the ultraviolet light in galaxies is more centrally concentrated than the HI, the UV/HI ratio (the Star Formation Efficiency) is nearly constant, with a slight dependency on surface brightness. This result is well accounted for in a model where disks maintain a constant stability parameter Q.  This model also accounts for how the ISM and star formation are distributed in the bright parts of galaxies, and how HI appears to trace the distribution of dark matter in galaxy outskirts.

\end{abstract}

\firstsection 
              
\section{Introduction}

The realization of the importance of the outskirts of galaxies as an environment for
disks to evolve and create new stars goes back to the beginning of
extragalactic \HI\ studies, 50-60 years ago, when the large
extent of galaxies in \HI\ was first noticed \cite[(e.g.\ van de Hulst \etal\ 1957;
H{\"o}gland \&\ Roberts 1965; Roberts 1966; Gordon \etal\
1968)]{vdHRvW1957,hr1965,roberts1966,grr1968}. Here I will emphasize
some studies that I found to be particularly inspiring in this field
mainly by astronomers attending this meeting.

Albert Bosma's doctoral research \cite[(Bosma 1981a,b)]{bosma1981a,bosma1981b} showed that at large radii $r$ the observed \HI\ surface density in spiral galaxies traces the projected total mass density as derived from rotation curves. Since this is dominated by Dark Matter (DM) in galaxy outskirts this result has typically been interpreted as \HI\ traces DM, and has been dubbed the ``Bosma effect'' \cite[(e.g.\ Carignan \etal\ 1989; Carignan \&\ Puche 1990a,b; Carignan \etal\ 1990; Hoekstra \etal\ 2001, Hessman \&\ Ziebart 2011)]{cb89,cp90a,cp90b,ccbv90,hvs01,hz11}. Explanations for it include that DM is gaseous \cite[(e.g.\ Pfenniger \etal\ 1994; Pfenniger \&\ Combes 1994)]{pcm94,pc94} or that there is no DM in galaxy outskirts, but rather the \HI\ disk dominates and that the gravitational force-law is given by Modified Newtonian Dynamics (MOND; \cite[Sanders \&\ McGaugh 2002]{sm02}). Either interpretation presents huge problems for the more accepted paradigm of the Cold Dark Matter (CDM) dominated galaxy evolution.  Under CDM dark matter should be dissipationless, whereas a gaseous baryonic DM would be dissipative, while the MOND scenario requires no DM.

Robert \cite[Kennicutt (1989)]{kennicutt89} demonstrated that the star
formation in spiral galaxies, as traced by \Halpha\ emission, often has
a sharp edge, beyond which there are few if any \HII\ regions.  He
showed that this edge typically corresponds to an increase in the
\cite[Toomre (1964)]{toomre64} stability parameter $Q$.  Thus the
formation of massive stars is limited to the dynamically unstable
portions of galaxies, while the dynamically stable outer disk is
virtually free of star formation. This result was reaffirmed using a
larger sample and better data by \cite[Martin \&\ Kennicutt
(2001)]{mk01}.

Since the work of \cite[Kennicutt (1989)]{kennicutt89} it became apparent that outer disks were not totally ``dead''.  The deep \Halpha\ images of Annette \cite[Ferguson (\etal\ 1998)]{fwgh98} showed that faint \HII\ regions were common beyond the star formation edge (also shown by \cite[Martin \&\ Kennicutt 2001]{mk01}).  With the launch of the GALEX satellite it became clear that outer disks were far from dormant when observed in a different star-formation tracer: ultraviolet (UV) emission. Spectacular examples of Extended Ultraviolet (XUV) disk galaxies were shown in papers lead by Dave \cite[Thilker (\etal\ 2005)]{thilker+05}, and Armando \cite[Gil de Paz \etal\ (2005)]{gildepaz+05}. The case of M83 is particularly troubling: the distinct edge seen in the \Halpha\ radial profiles of \cite[Martin \&\ Kennicutt (2001)]{mk01} were not at all apparent in far UV profiles (\cite[Thilker \etal\ 2005]{thilker+05}). 

Below, I present some results from projects I have been doing primarily with students and postdocs on gas and star formation in the outer disks of galaxies. After describing our star formation surveys in Sec.~\ref{s:ss}, I summarize our ongoing work on surveying star formation in the outskirts of galaxies, in both \Halpha\ and UV emission in Sec.~\ref{s:survod}. In Sec.~\ref{s:mod} I present a model for explaining the distribution of gas and star formation in galaxies, which is particularly relevant for outer disks.

\begin{figure}[hb]
\begin{center}
 \centerline{\includegraphics[height=6.0cm]{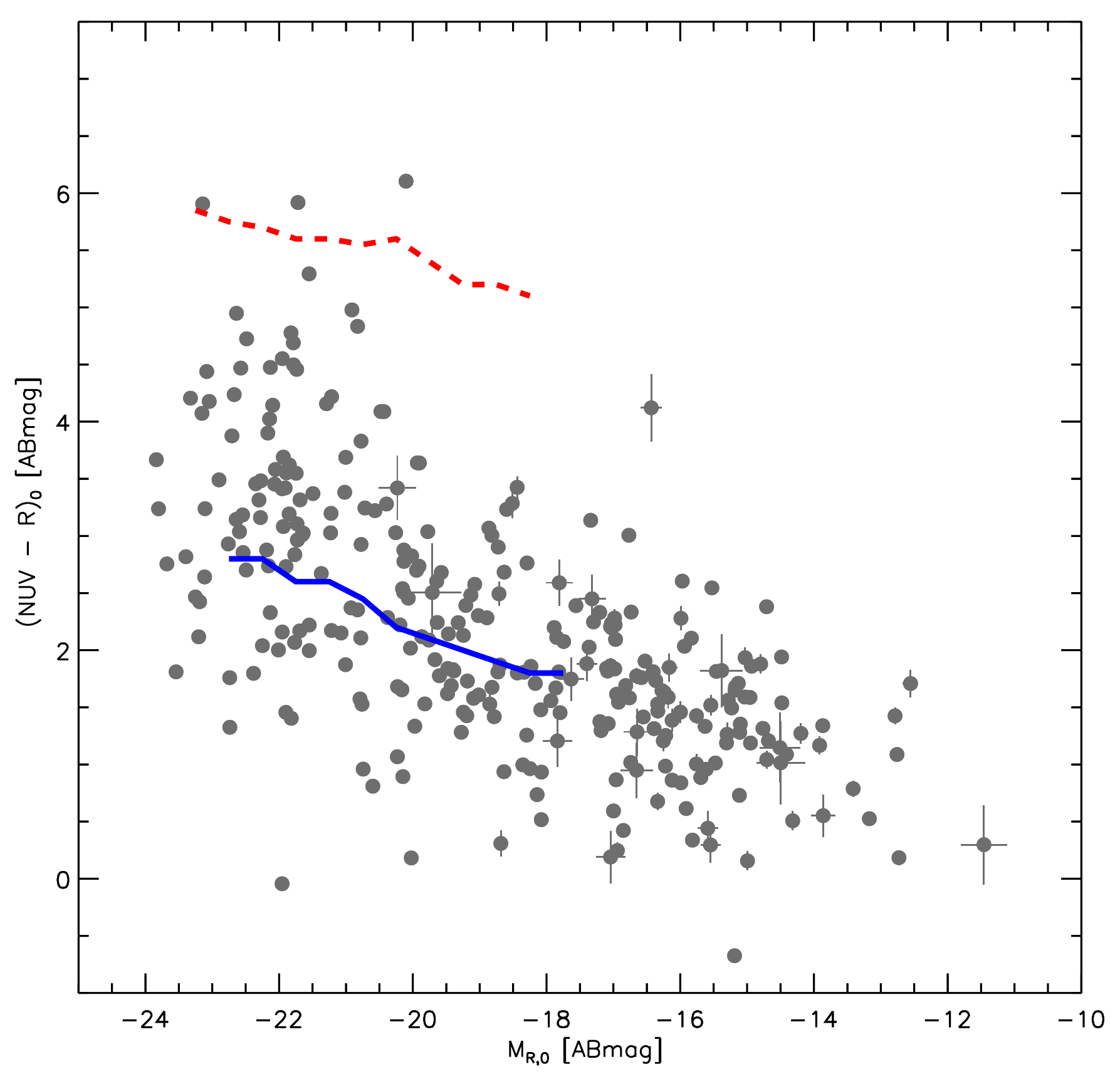} 
 \includegraphics[height=6.0cm]{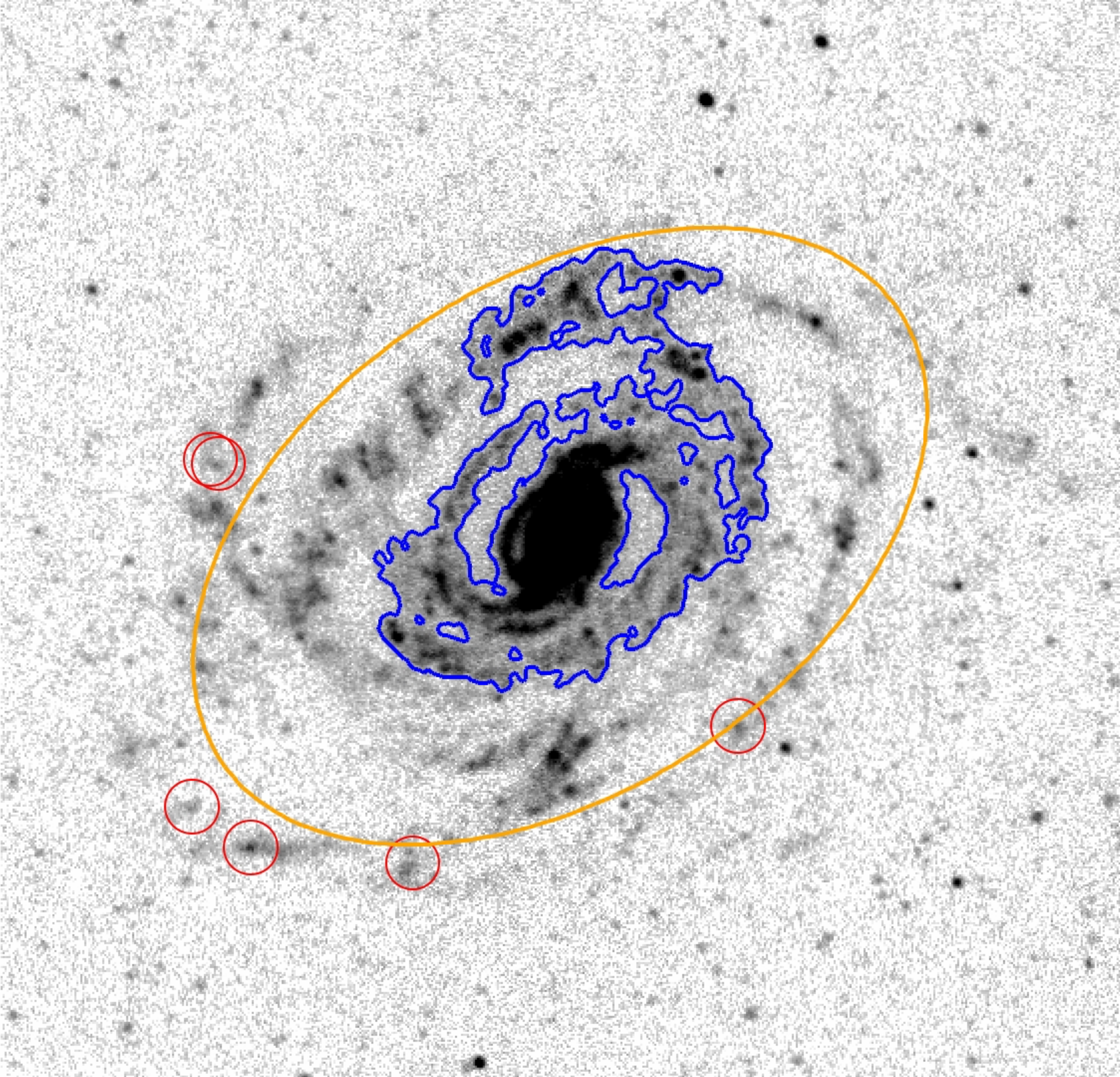}} 
\caption{{\bf (a) Left}: Ultraviolet - optical dust absorption
  corrected color-magnitude diagram of galaxies in both the SINGG and
  SUNGG surveys. The dashed and solid lines mark the ridge lines for
  the passive and star-forming sequences, respectively, of galaxies
  selected from the SDSS \cite[(Wyder \etal\ 2007)]{wyder+07}.  The
  latter is often known as the ``main-sequence of star forming
  galaxies''. Note how the star forming sequence can be traced to much
  lower luminosities in our surveys and that there are very few \HI\
  selected passive sequence galaxies. {\bf (b) Right}: GALEX NUV image
  of HIPASS J0052--31 (NGC~289). The irregular contour corresponds to
  the NUV surface brightness $27.35\, {\rm ABmag\, arcsec^{-2}}$ used
  to identify XUV1 disks. The extensive UV emission beyond this
  isophote marks J0052-31 as an XUV1 disk galaxy. The ellipse shows
  the area beyond which \HII\ regions can be classified as ELdots. Six
  \Halpha\ sources classified as ELdots are circled. }.
   \label{f:cmd_examp}
\end{center}
\end{figure}
 
\section{SINGG and SUNGG}\label{s:ss}

Our work is primarily based on an \HI\ selected sample of galaxies
studied in the light of two different star formation tracers.  The
starting point is the \HI\ Parkes All Sky Survey (\HIPASS\ \cite[Meyer
\etal\ 2004; Koribalski \etal\ 2004; Wong \etal\ 2006]{meyer+04,koribalski+04,wong+06nhicat}).
\HIPASS\ found 4315 \HI\ sources south of a declination of
$\delta = +2^\circ$ with the Parkes 64m radio telescope. We selected a sample of 468 \HIPASS\ sources to
observe in \Halpha\ and the $R$ band continuum for the Survey for
Ionization in Neutral Gas Galaxies (SINGG) primarily with the CTIO
1.5m and 0.9m telescopes \cite[(Meurer \etal\ 2006)]{meurer+06}.  A
total of 331 \HIPASS\ targets were observed, often revealing multiple
\Halpha\ emitting galaxies.  Similarly we selected 139 \HIPASS\
targets from the same parent sample to observe in the far and near UV
(FUV and NUV) for the Survey for Ultraviolet emission in Neutral Gass
Galaxies (SUNGG; \cite[Wong 2007]{wong07}). Using archival data and also
analyzing other \HIPASS\ sources that lie in the same GALEX field of
view, we ended with observations of 313 \HIPASS\ targets.

No survey is without biases.  Ours are biased towards the star forming
mains sequence, and away from the red sequence, as shown in
Figure~\ref{f:cmd_examp}a. This is expected.  Our surveys are meant to
target galaxies with an interstellar medium (ISM) and hence those that
are capable of forming stars, and then search for new stars in them.
Since the red sequence is predominantly free of a cool or cold ISM
very few are in our parent sample. A strength of our selection is that
we can trace star forming galaxies much further down the star forming
main-sequence than traditional flux limited surveys, as shown in the
Figure.  We are still able to derive properties averaged over large
cosmic volumes by tying our results to the \HI\ mass function as done
by \cite[Hanish \etal\ (2006)]{hanish+06} who calculated the \Halpha\ based star
formation rate density of the local universe from SINGG.

\begin{figure}[hb]
\begin{center}
\centerline{\includegraphics[width=14.0cm]{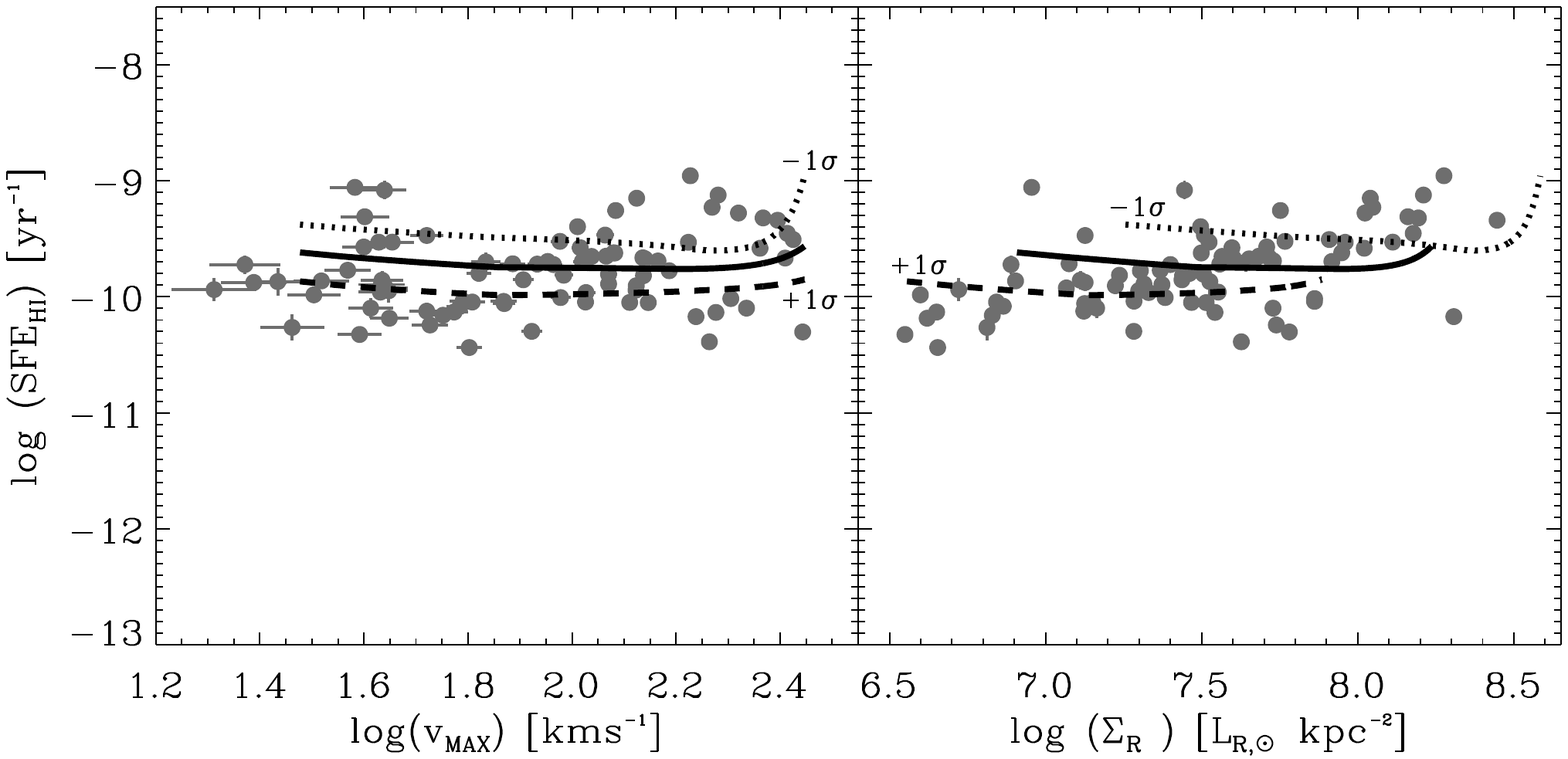}}
 \caption{Star Formation Efficiency as a function of circular velocity
   \Vmax\ (left) and effective $R$-band surface brightness (right).
   Data from SUNGG are shown as circles. Our standard model sequence
   is shown with the solid line. The dotted (dashed) line show cases
   where the spatial scale for each model was contracted (expanded) by
   a factor of 1.4 to simulate a decrease (increase) in angular
   momentum. For full details see \cite[Wong \etal\ (2016)]{wong+16}.}
   \label{f:sfeuv}
\end{center}
\end{figure}

The star formation tracers \Halpha\ and FUV are sensitive to O stars, and both O and B stars, respectively.  The lower limit to the initial masses of these stars are 18 and 3 \Msun, respectively.  Hence, comparing results from these surveys allows us to probe the Initial Mass Function (IMF) of stars, although ones has to be careful to account for other parameters that effect this ratio, specifically star formation history, and dust content. Probing the IMF was one of the major goals of our surveys, and we did so in \cite[Meurer \etal\ (2009)]{meurer+09}. There we showed that the ratio of \Halpha\ to FUV fluxes varies systematically with several parameters, most strongly with $R$-band surface brightness. Similar results were found by Janice \cite[Lee \etal\ (2009)]{lee+09hafuv}.  After carefully considering other possibilities we argued these results are most easily explained with an IMF deficient in high mass stars in low surface brightness galaxies \cite[(Meurer \etal\ 2009)]{meurer+09}.

\section{Surveying outer disk star formation}\label{s:survod}

\underline{\it Surveying with SINGG: ELdots.}  We searched the SINGG dataset for small line emitting sources projected far from any obvious host.  We dubbed Emission Line Dots or ELdots (\cite[Ryan-Weber \etal\ 2004; Werk \etal\ 2008; Werk \etal\ 2010a,b; Werk \etal\ 2011; Santiago-Figueroa \etal\ 2011]{ryan-weber+04,werk+08,werk+10a,werk+10b,wpms11,spwmr11}).  The name perfectly describes their apparent properties, but leaves open their nature. The ambiguity is essential since without additional information it is unclear where these sources are. They may be \Halpha\ from an outer \HII\ region, a tiny dwarf galaxy in the immediate vicinity of the \HI\ target, a small galaxy up to several Mpc away (see e.g. Santiago-Figueroa \etal\ 2011) or a bright background galaxy emitting in \fion{O}{III}, \fion{O}{II} or some other bright line. While, in spectacular cases many ELdots are arranged in obvious tidal arms indicating most are local to the host, spectroscopy or very high resolution imaging (e.g.\ HST; \cite[Werk \etal\ 2008]{werk+08}) are the only sure ways to determine where the ELdots are relative to the \HIPASS\ target. However, combining UV and \Halpha\ photometry allows likely background ELdots to be identified by their relatively red $FUV-R$ color (\cite[Werk \etal\ 2010a]{werk+10a}).

Systems hosting local ELdots often show signs of interaction or
mergers (\cite[Ryan-Weber \etal\ 2004; Werk \etal\
2010a]{ryan-weber+04,werk+10a}). \cite[Werk \etal\ (2011)]{wpms11}
searched for ELdots around merging systems from the ``\HI\ Rogues''
sample (\cite[Hibbard \etal\ 2001]{hibbard01}), and determined their
metallicities spectroscopically to test the hypothesis that these
outer \HII\ regions may represent the first star formation in 
chemically pristine ISM from the outskirts of galaxies.  However, they
found the metallicity gradients are weak and the merger outskirts have
uniform modest abundances (typically around half
solar). This is also true for the cases of ELdots in systems
that are isolated like NGC~2915 (\cite[Werk \etal\ 2010b]{werk+10b}) which consists of a blue compact dwarf galaxy surrounded by a spectacular \HI\ disk \cite[(Meurer \etal\ 1994; 1996)]{mmc94,mcbf96}.

We have identified 219 ELdots in 104 \HIPASS\ targets in the full
SINGG dataset; about one third of all the \HIPASS\ fields we surveyed contain at least one ELdot 
(El Agali \etal\ 2016 in prep). We have deep GALEX data (exposure
times greater than 1000s), for 36 of the fields with ELdots allowing a
crude classification of ELdots as local or background. Of the 106
ELdots in those fields, slightly more than half, 58, were found to be
local (most likely outer \HII\ regions although see Santiago-Figueroa \etal\ 2011). A large
fraction ($\sim$1/3) of the hosts appear to be strongly interacting, while nearly
half have at least one companion.

\underline{\it Surveying with SUNGG: XUV disk galaxies.} Over the austral summer 2015/6 we began a project to classify the SUNGG GALEX images in search of extended UV (XUV) disks using the definitions of \cite[Thilker \etal\ (2007)]{thilker+07}.  We concentrated on type 1 XUV disks (XUV1), identified by having obvious signs of star formation beyond an UV isophote designed to be similar to the star formation threshold of \cite[Kennicutt (1989)]{kennicutt89}.  This initial morphological classification was done by summer student Leah Kalimeris, and were checked by Dave Thilker and myself.

We started with 360 \HIPASS\ Targets containing 398 galaxies (i.e.\
including multiples from SINGG). After rejecting defective images, and other problems, we were left with 332 galaxies to
classify. We found 23\%\ are clearly XUV1; 31\%\ clearly are not, and
46\%\ were uncertain. Note these numbers are the preliminary values
presented at the meeting. I can confidently say the final fraction of
XUV1 galaxies in our sample will be higher than 23\%\ but will depend
on how we draw the line between what is clearly an XUV1 galaxy and
what isn't (Kalimeris \etal\ 2016, in prep.).

In comparison, \cite[Thilker \etal\ (2007)]{thilker+07} considered a
sample of 189 galaxies within 40 Mpc from the GALEX Nearby Galaxy
Atlas (\cite[Gil de Paz \etal\ 2007]{gildepaz+07}) having  optical
morphologies between S0 and Sm and found 16\%\ to be XUV1 disks. The
higher rate of XUV1 galaxies in our (considerably larger) sample may
be due to the \HI\ selection which includes many more dwarf galaxies
and avoids gas poor disks.

\underline{\it Overall Impressions.}  The definitions for ELdots and XUV disks differ in detail but both select star formation beyond the typically defined outer edge of galaxies. With both methods we find $\gapeq 20$\%\ of \HI\ selected galaxies have outer disk star formation. There are many cases of galaxies with both ELdots and XUV1 disks, an example is shown in Figure~\ref{f:cmd_examp}b. The overall impression I get is that XUV disks are more common than galaxies with ELdots, and similarly that outer disks are more apparent in the UV than \Halpha. This impression hints at the possibility that the IMF in the outer disk is deficient in high mass stars as seen in LSB galaxies (\cite[Meurer \etal\ 2009]{meurer+09}).  This is at apparent odds with the results of \cite[Koda \etal\ (2012)]{koda+12} who find that the statistics of the number of UV bright knots in the outer disk of M83 with and without \Halpha\ emission as well as the integrated \Halpha\ to FUV flux ratio from these knots are consistent with a standard IMF, that is like the bright portion of galaxies. However, in that study only the light in these knots are measured; their method is insensitive to diffusely distributed stars. To square this result with the sharp difference between the integrated \Halpha\ and FUV surface brightness profiles of M83 noted in the introduction requires that there be considerable UV bright star formation in lower mass clusters or in individual stars.  This is consistent with HST imaging of the outer disks of NGC~2915 (\cite[Bruzzese \etal\ 2015]{bruzzese+15}) and M83 (\cite[Bruzzese 2016]{bruzzese16}) where the obvious young clusters contain only a small portion of the main-sequence stars. Further work is clearly needed to understand the nature of low intensity star formation in both dwarf galaxies and outer disks.

\section{Disks and Galaxy Evolution}\label{s:mod}

We are developing a model to explain the range of
properties seen along the main sequence of star forming
galaxies. Inspired by \cite[Leroy \etal\
(2008)]{leroy+08}, as well as earlier studies (\cite[Kennicutt
1989]{kennicutt89}; \cite[Meurer \etal\ 1996]{mcbf96}; \cite[Martin
\&\ Kennicutt 2001]{mk01}), it became clear that disk stability plays
an important role in determining disk structure. Hence, the underlying
assumption of our model is that galaxy disks evolve to have a uniform
stability.

In \cite[Meurer \etal\ (2013)]{mzd13} we demonstrated that in the
limit of a flat rotation curve, the gas density in a constant $Q$ disk
falls off with radius as $r^{-1}$, the same projected mass
distribution of the DM halo.  This then is the explanation for the
Bosma effect - disks equilibrate to trace the dominant mass when
the rotation curve is flat.  It also accounts for the
anti-correlation of the ratio of \HI\ and total projected mass density
with rotational amplitude, a result that neither the gaseous DM
hypothesis nor MOND can explain. 

In \cite[Zheng \etal\ (2013)]{zheng+13} we turned our attention to modelling the optically bright parts of galaxies. Here both gas and stars are important to the mass budget of disks so a two fluid version of the $Q$ parameter is required. We used data from the THINGS survey (\cite[Leroy \etal\ 2008]{leroy+08}) to construct our models, which require the rotation curve and the stellar mass distribution as inputs. From the definition of the two fluid $Q$, we determine the distribution of the ISM disk, which is apportioned in to molecular and neutral phases. A star formation law then allows us to determine the resulting distribution of star formation. We found that our model roughly reproduces the relevant observed profiles but typically do not match them in great detail.  The center can be particularly challenging as the steeply rising rotation curves of bulge dominated galaxies require a much larger central gas content and star formation intensity than usually seen.

We next turned our attention to a curious result found by several
teams, and verified by us - that the galaxy averaged ratio of FUV
emission to 21cm \HI\ emission is fairly constant over a wide range of
parameters \cite[(Schiminovich \etal\ 2010; Huang \etal\ 2012; Wong
\etal\ 2016)]{schiminovich+10,hhgb12,wong+16}. This ratio is often
referred to as the Star Formation Efficiency, SFE, the ratio of
star formation rate to gas content. I use the term SFE for
convenience, but one must bear in mind the standard caveats, namely
that UV fluxes are highly susceptible to dust absorption, while high
mass star formation correlates more strongly with the molecular rather
the neutral component of the ISM in the centers of galaxies (e.g.\
\cite[Bigiel \etal\ 2008]{bigiel+08}).  On the other hand, FUV
emission is a direct tracer of star formation, unlike \Halpha\ and far
infrared emission. It is also less prone to uncertainties in the upper
end of the IMF than \Halpha\ emission (\cite[Meurer
\etal\ 2009; Lee \etal\ 2009]{meurer+09,lee+09hafuv}). The ISM in galaxy 
outskirts is dominated by \HI\ while the molecular ISM is
notoriously difficult to detect.  Meanwhile, UV emission and blue
stars correlate well with \HI\ emission there (\cite[Cuillandre \etal\
2001; de Blok \&\ Walter, 2003; Bigiel \etal\ 2010a,b; Bruzzese \etal\
2015]{cuillandre+01,dBW03,bigiel+10a,bigiel+10b,bruzzese+15}).

The insensitivity of SFE to galaxy properties is puzzling because star
formation in galaxies is much more centrally concentrated than \HI\
emission \cite[(e.g.\ Leroy \etal\ 2008)]{leroy+08}.  Hence, how does the outer \HI\ ``know'' how
to balance the centralized star formation?

In \cite[Wong \etal\ (2016)]{wong+16} we account for this relationship
using our constant $Q$ model. The driving property of the model - halo
mass, is parameterized by the peak orbital velocity \Vmax. The shape
of the rotation curve for a given \Vmax\ is taken from the
parameterization of \cite[Persic \etal\ (1996)]{pss96}. We use observed
scaling relations to set the stellar mass distribution which we put in
a pure exponential disk \cite[(Freeman 1970)]{freeman70}. We adopt the
\cite[Wang \&\ Silk (1994)]{ws94} formula for the two fluid $Q$, and find the best results when we apportion the neutral and
molecular phases using a recipe based on the disk mid-plane hydrostatic
pressure.  As noted above, the radial distribution of the outer disk
declines as $r^{-1}$, producing an infinite integral if a cutoff is
not introduced.  Analysis of data from SINGG, SUNGG, and the sample
used by \cite[Meurer \etal\ (2013)]{mzd13} indicates the outer
edge of the disk in the optical, UV and \HI\ are all given by the
radius where the orbital time is $\sim$ 1 Gyr (Meurer \etal\ 2016, in
prep.). We adopt this cutoff in our model.  After adopting a molecular
star formation law, we have radial profiles of \HI\ and SFR which we
integrate to determine SFE.

Figure~\ref{f:sfeuv} shows our model fits the flat SFE relationship
with \Vmax\ very well. We find a slight correlation between SFE and R
band effective surface brightness.  We interpret this as due to
angular momentum differences between galaxies. This make sense in that
after mass, angular momentum of the halo is the second most important
parameter in setting the structure of galactic halos and hence the
galaxies they contain \cite[(Mo, Mao, \&\ White 1998)]{mmw98}.  To
model the effects of a change in the angular momentum we contract or
expand the spatial scale of our standard model by a factor of 1.4, 
the amount required to match the scatter in the luminosity -
surface brightness correlation.  This produces the offset lines shown in the
figure. We see that the offset curves account reasonably well for the
scatter in SFE versus \Vmax, and the tilt of the SFE versus
$\Sigma_\star$ data.

\section{Summary}

\HI\ selected samples of galaxies are ideally suited to understanding
the nature of the outer disks of galaxies because gas in galaxies
naturally settles into disks, and the \HI\ survives best in the low
density low pressure outskirts of disks.  Using \Halpha\ and UV
follow-up observations of the \HIPASS\ survey we find over 20\%\ of
\HI-selected galaxies have star formation in their outskirts as seen
by \HII\ regions or Type 1 XUV disks. We developed a model for
the distribution of star formation and the gas phases based on the
premise that galaxies evolve to have a constant stability parameter
$Q$. We showed that this model accounts for (a) the ``Bosma
effect'', whereby \HI\ appears to trace the distribution of Dark
Matter in the outskirts of galaxies; (b) predicts (albeit crudely)
the radial distribution of star-formation and the ISM phases
(molecular and neutral) in the optically bright portion of galaxies,
and (c) predicts the lack of correlation of the galaxy averaged Star
Formation Efficiency (based on \HI\ and UV measurements) with most
global properties of galaxies.

{\bf Acknowledgments} I gratefully acknowledge the students and postdocs who lead much of the work I reported here: Fiona Audcent-Ross, Ahmed Elagali, Leah Kalimeris, Ivy Wong, and Zheng Zheng.  I thank Dave Thilker for co-supervision of students and collaborating on many projects on the outer disks of galaxies. I thank the editors for their patience and a suggestion that improved this contribution.

\end{document}